\documentclass{kluwer}    

\newdisplay{guess}{Conjecture}
\usepackage{graphicx}

\begin{document}
\begin{article}
\begin{opening}         
\title{Maser action in methanol transitions}
\runningauthor{M.A. Voronkov et al.}
\runningtitle{Maser action in methanol transitions}
\author{Maxim \surname{Voronkov}}
\institute{Australia Telescope National Facility, Australia\\
Astro Space Centre, Russia}
\author{Andrej \surname{Sobolev}}
\institute{Ural State University, Russia}
\author{Simon \surname{Ellingsen}}
\institute{University of Tasmania, Australia}
\author{Andrei \surname{Ostrovskii}}
\institute{Ural State University, Russia}
\author{Alexei \surname{Alakoz}}
\institute{Astro Space Centre, Russia}

\date{April 15, 2004}

\begin{abstract}
  We report the detection with the ATCA of 6.7~GHz methanol emission
  towards OMC-1.  The source has a size between 40$''$ and 90$''$, is
  located to the south-east of Ori-KL and may coincide in position
  with the 25~GHz masers. The source may be an example of an interesting case
  recently predicted in theory where the transitions of traditionally
  different methanol maser classes show maser activity simultaneously.
  In addition, results of recent search for methanol masers
  from the 25 and 104.3~GHz transitions are reported.
\end{abstract}
\keywords{masers -- ISM: molecules}

\end{opening}           

\section{Introduction}
Interstellar methanol masers are associated with regions of active
star formation and masing transitions are traditionally divided into
two classes \cite{men}. Although masers of both classes often co-exist
in the same star-forming region, Class~I masers are normally seen
apart from the strong continuum sources, while Class~II masers are
found close to them.  The $5_1-6_0$~A$^+$ transition at 6.7~GHz (a
Class~II transition) produces the brightest known methanol maser (up
to 5000~Jy).  Maser emission from this transition is widespread and to
date has been detected towards more than 400 sources.  Among the
Class~I methanol masers there is the J$_2-$J$_1$~E series near 25 GHz.
The strongest $5_2-5_1$~E 25 GHz maser (about 150 Jy) is observed
towards the Orion molecular cloud (OMC-1) which is the best studied
Class~I source.  The 25 GHz transitions are seen in absorption
towards the archetypal Class~II source W3(OH) \cite{men86} and the
general assumption has been that other Class~II sources will exhibit
similar behaviour. To date only a
few 25~GHz methanol masers have been detected and bright 25 GHz masers
are thought to be rare (e.g., \opencite{men86}).  However, the survey
of \inlinecite{men86} did not cover many sites which subsequent searches in
other transitions have been found to be prominent Class~I methanol
maser sources. In this paper we report
results of a pilot survey of 25~GHz masers targeted at Class~I masers
at 44~GHz with sufficiently well known positions.

\begin{figure}
\centerline{\includegraphics[width=\linewidth]{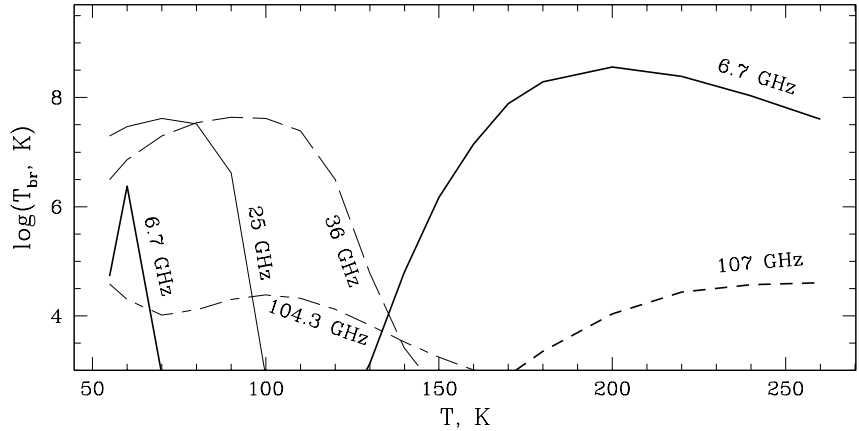}}
\caption{The model dependence of the brightness temperature of various
methanol
maser transitions on the temperature ($T=T_{gas}=T_{dust}$).}
\label{model}
\end{figure}

The difference between the two traditional classes is due to a
difference in the pumping mechanism: strong Class~I masers appear when
the excitation processes are dominated by collisions while the strong
Class~II masers appear when the radiative excitation prevails (e.g.,
\opencite{cra}). Therefore, the masers in transitions of different
classes are usually seen apart from each other.  The work of \inlinecite{cra}
was limited to the case where conditions in the masing
region are such that the radiative transitions between different
torsionally excited states are not influential.

Further studies of methanol maser excitation have shown that the
strong Class~II masers are produced when excitation from the ground to
the second torsionally excited state comes into effect (see, e.g.,
\opencite{sob94}; \opencite{sob97a}).  Calculations show
that the involvement of torsional transitions does not allow masers of
different traditional classes to become strong simultaneously.
However, the coexistence of inversion in some Class~I and II transitions
is predicted in some situations  (e.g., the Class I $11_2-11_1$~E transition is
present in the list of Class II methanol maser candidates of \opencite{sob97b}).

The general tendencies of methanol maser pumping are demonstrated in
Fig.\ref{model} showing the dependence of brightness temperature in
different masing (i.e., inverted) transitions on the temperature. In
this model the collisional excitation is provided by the hydrogen
molecules and the pumping radiation is produced by the dust intermixed
with the gas within the cloud. The cloud has the following parameters:
hydrogen number density $10^5~cm^{-3}$, methanol abundance $10^{-5}$,
beaming factor ($\varepsilon^{-1}$) 10, specific (divided by
linewidth) column density $10^{11.5}$ cm$^{-3}$s, temperature of the
dust and gas are equal.  Increasing the temperature in such a cloud
changes the balance of excitation processes in favour of radiative
ones.  So, one should expect that the Class~I masers should appear at
low temperatures while at high temperatures the Class~II masers should
shine brightly.  The figure shows that this is correct, at least for
well known Class~I masers at 25~GHz and 36~GHz and Class~II maser at
107~GHz. In addition, the figure shows that there is some transition
region when the strong Class~I $5_2-5_1$~E maser can coexist with the
weaker maser in Class~II $5_1-6_0$~A$^+$ transition.  This is not
expected in the traditional classification scheme and we have
attempted to test this prediction through observation. Further,
Fig.~\ref{model} shows that the maser at 104.3~GHz, which was
predicted in theory \cite{vor}, could be a representative of
intermediate class. If the maser appears mostly under intermediate
conditions then the detection rate in surveys targeted towards either
Class~I or Class~II sources may be low.

\section{Observational Results and Discussion}
\subsection{Search for 6.7~GHz emission from the 25 GHz maser site in OMC-1}
\begin{figure}
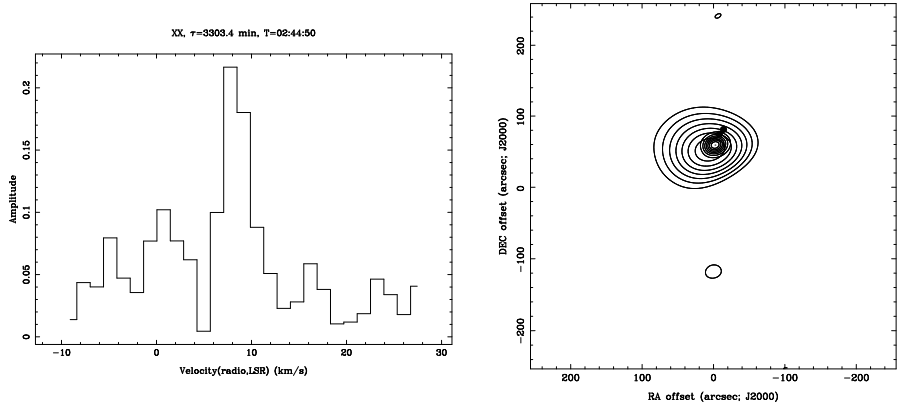

\centerline{%
\parbox{0.5\linewidth}{\rotatebox{270}{\includegraphics[height=\linewidth]{line6all.ps}}}
\hfill\parbox{0.45\linewidth}{\rotatebox{270}{\includegraphics[width=\linewidth]{6and25lines.ps}}}}
\caption{Left. The spectrum of detected 6.7~GHz emission in OMC-1. Amplitude is in Jy.
Right. The combined map at 6.7~GHz and 25~GHz (a peak spectral channel
corresponding to 8 km~s$^{-1}$ was used for imaging). Two filled circles represent the positions of BN and KL
objects.}
\label{orion6and25}
\end{figure}

The Australia Telescope Compact Array (ATCA) was used to search for 6.7~GHz
emission in OMC-1. Observations were made with the array in the compact
H75 configuration on 10 June 2003.
A weak spectral feature (about 0.2~Jy) has been detected
(Fig.\ref{orion6and25}, left). Using the newly installed 12-mm system
at the ATCA, the 25~GHz maser was observed quasi-simultaneously with
the 6.7~GHz observations. For both frequencies the correlator was
configured to split the 4~MHz bandwidth into 1024 spectral channels.
The spectral resolution at 25~GHz was 0.047~km~s$^{-1}$.
At 6.7~GHz each 8 adjacent channels were averaged together reducing the
effective spectral resolution to 1.4~km~s$^{-1}$. 
Because the spatial resolution of these
observations at 6.7~GHz is rather coarse we have compared that data
with our 25~GHz measurement, although images with a better spatial
resolution exist for 25~GHz maser in OMC-1 \cite{joh}.  In
Fig.\ref{orion6and25}(right) the combined image at the two frequencies
is shown. A peak spectral channel corresponding to
$V_\mathrm{LSR}=8.0$~km~s$^{-1}$ was used for imaging at both
frequencies.  It follows from this image that the detected 6.7~GHz
emission may come from the same region of space as the 25~GHz maser
emission. The peak pixels in the 6.7~GHz and 25~GHz images have
offsets from the phase centre ($\alpha_{2000}=5^h35^m15^s$,
$\delta_{2000}=-5^o23'43''$) equal to (5, 50) and ($-$11, 67) arcseconds
respectively. This gives a displacement of about 23''. The estimate
of the displacement $3\sigma$  accuracy is about 60'' (the synthesized beam
size is about 2' and the signal to noise ratio is about 3 at 6.7~GHz).
The peak of the 25~GHz emission tends to move towards that at 6.7 GHz
if a spectral channel corresponding to lower velocity is used for
imaging. A minimum separation of about 10'' achieved
at $V_\mathrm{LSR}=7.3$~km~s$^{-1}$.
This behavior is in agreement with the 25~GHz map of \inlinecite{joh}, where
the maser has been resolved into a number of spots at close velocities
spread across the region of about 30'' in extent. Due to a poor spatial
resolution of the ATCA measurement the position we measure is a weighted
(with flux) averaged position of individual maser spots.
A higher resolution study is
desirable to measure the position of the 6.7~GHz emission with respect
to individual 25~GHz maser spots (say from the map of \opencite{joh}).
However, it is difficult to do this with existing interferometers
because the source is very weak at 6.7~GHz and possibly resolved.

On the basis of these observations we cannot prove that the emission
detected at 6.7~GHz is a maser (brightness temperature $T_b>1$~K).
However, this is not in contradiction
with the theory. Fig.\ref{model} shows that the bright 25~GHz masers
are not necessarily accompanied by the maser emission at 6.7~GHz. So,
the actual physical parameters (density, temperature, etc) of the
source may be different from the interesting pumping regime where the
two maser transitions of different classes do coexist.  If the
emission is quasi-thermal its detection towards the Class~I maser
position in OMC-1 is interesting because according to traditional
reasoning the Class~II transitions at such locations should be
overcooled (i.e. the upper level population should be less than the
equilibrium value). So, it is expected that the corresponding lines will be
undetectably weak or in absorption.  The source appears to be
unresolved on all short baselines, although it was not detected on the
baselines including the 6km antenna. We reobserved the source at
6.7~GHz using the 1.5D configuration, which contains a greater number
of large baselines.  These observations can be used to place limits on
the brightness temperature at 6.7~GHz, which are helpful for
understanding the nature of the detected emission. However, the signal
was detected on the shortest baseline only, which implies the source
size less than about 90'' and greater than about 40'' (0.6~K$<T_b<$3.4~K).
It is worth noting that an extended source can indicate either a thermal or
maser nature, while a compact source must be a maser. 

\subsection{Search for new bright 25 GHz masers}
\begin{figure}
\centerline{\includegraphics[width=\linewidth]{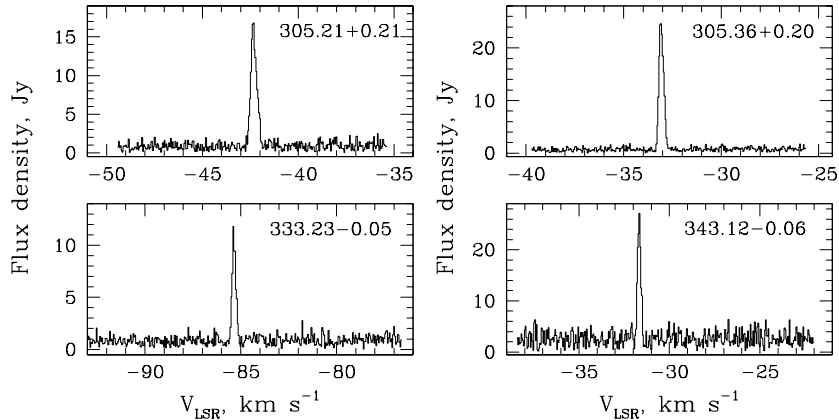}}
\caption{New 25 GHz masers. The phase was self-calibrated and all baselines
were averaged together decreasing the noise level in the spectra.}
\label{new25}
\end{figure}

The southern sky is very attractive for Class~I maser searches because
the most widespread Class~I methanol maser line at 44~GHz has been
searched for in southern sources \cite{sly} giving a good target list.
The ATCA in the 1.5D configuration was used to search for new
$5_2-5_1$~E 25~GHz masers. The resulting detection limit was about
5~Jy and we observed the following 10 sources: 305.21$+$0.21,
305.36$+$0.20, 333.23$-$0.05, 343.12$-$0.06, M8E, 333.13$-$0.43,
335.59$-$0.29, 337.91$-$0.47, 338.92$+$0.56, 351.78$-$0.54. The first
4 sources listed are new detections (Fig.\ref{new25}).
The results of this pilot survey mean that the
25~GHz masers may be rather common. There is no simple correlation
between the flux densities at 25~GHz and 44~GHz. Interestingly, the
brightest known 44~GHz maser in M8E showed no emission in the 25~GHz
transition.

\subsection{The 104.3~GHz maser search}
\begin{figure}
\centerline{\includegraphics[width=0.5\linewidth]{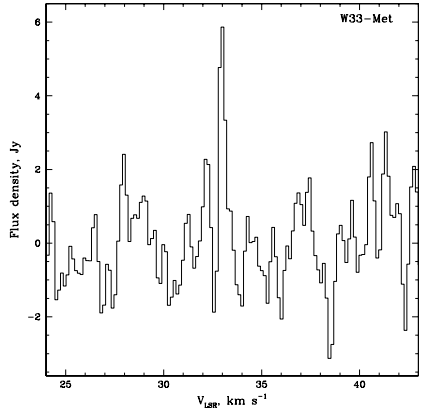}}
\caption{The 104.3~GHz spectrum of W33-Met.}
\label{w33met}
\end{figure}
\begin{figure}
\centerline{\includegraphics[width=\linewidth]{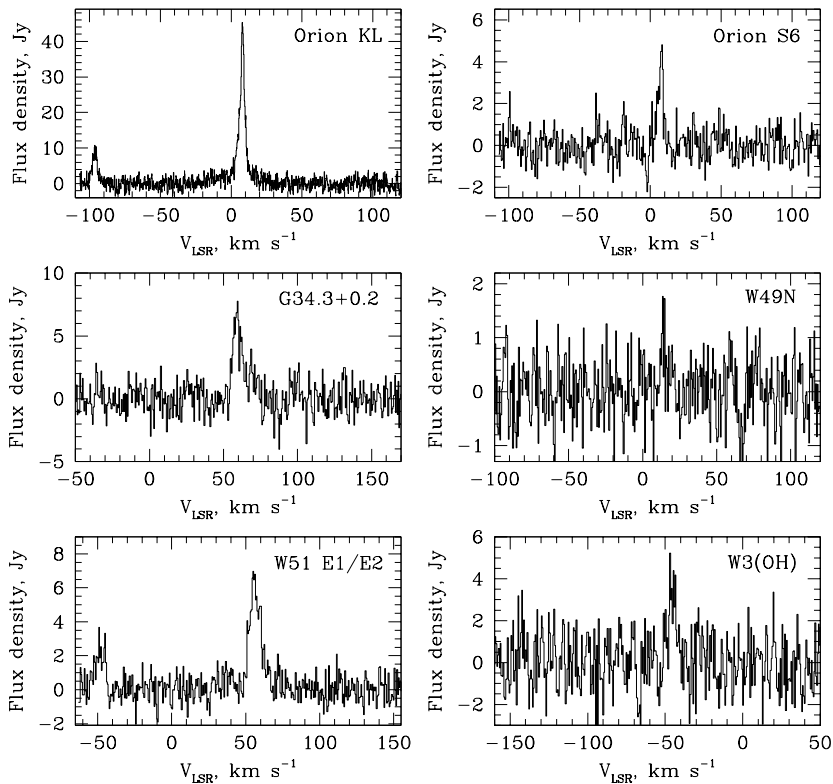}}
\caption{Sources with a broad line emission detected at 104.3~GHz.
A weaker feature which is present in the left hand side of some spectra
belongs to CH$_3$OD (the same transition).}
\label{broad104}
\end{figure}

A search for new methanol masers in the $11_{-1}-10_{-2}$~E transition
at 104.3~GHz was carried out with the Onsala radiotelescope. The maser
in this transition was predicted in theory \cite{vor} and can be
bright at Class~I maser positions. Our target list included all known
44~GHz masers visible from the high latitude of the Onsala
observatory. In addition to them two Class~II maser sources, W3(OH)
and GL2789, were also observed as they can have a thermal emission and
they were above the horizon almost all the time.  The only obvious
maser (i.e. a source with a very narrow line) detected was W33-Met
(Fig.\ref{w33met}).  In addition to this source, 6 sources showed
broad line emission, which is probably thermal (Fig.\ref{broad104}).
\inlinecite{sly93} searched for the $9_{-1}-8_{-2}$~E masers at
9.9~GHz. This maser transition belongs to the same series
$(\mathrm{J}+1)_{-1}-\mathrm{J}_{-2}$~E and should have a similar
behavior. \inlinecite{sly93} have found 9.9 GHz maser also towards
W33-Met only. It appears that, the masers at both 9.9~GHz and
104.3~GHz are rare and may be representatives of an intermediate
class of methanol masers.

\section{Conclusions}
\begin{enumerate}
\item 6.7~GHz emission which may be associated with the 25~GHz maser
has been detected towards OMC-1. The source has a size between 40'' and 90''.
\item New 25 GHz methanol masers were discovered in 305.21+0.21,
305.36+0.20, 333.23-0.05, 343.12-0.06.
\item 104.3 GHz methanol emission has been detected towards Orion KL,
Orion S6, G34.3+0.2, W49N, W51 E1/E2, W3(OH) and W33-Met. In the
last source the emission is a maser.
\end{enumerate}

\begin{acknowledgements}
We would like to thank the local staff of Narrabri and Onsala observatories
for the help during observations. The Australia Telescope is funded by
the Commonwealth of Australia for operation as a National Facility managed
by CSIRO. Maxim Voronkov and Alexei Alakoz
were partially supported by the RFBR grant no. 01-02-16902 and
by the program "Extended objects in the Universe-2003". Andrej Sobolev and
Andrei Ostrovskii were supported by the RFBR grant no. 03-02-16433.
\end{acknowledgements}

\end{article}
\end{document}